\documentclass[sigconf]{acmart}
\AtBeginDocument{%
  }

\setcopyright{acmlicensed}
\copyrightyear{2018}
\acmYear{2018}
\acmDOI{XXXXXXX.XXXXXXX}
\acmConference[Conference acronym 'XX]{Make sure to enter the correct
  conference title from your rights confirmation email}{June 03--05,
  2018}{Woodstock, NY}
\acmISBN{978-1-4503-XXXX-X/2018/06}

\begin{document}

\title{Navigating the Future of Federated Recommendation Systems with Foundation Models}


\author{Zhiwei Li}
\orcid{0000-0002-8897-8905}
\affiliation{%
  \institution{Australian AI Institute, Faculty of Engineering and IT, University of Technology Sydney}
  \city{Sydney}
  \state{NSW}
  \country{Australia}
}
\email{zhiwei.li@student.uts.edu.au}

\author{Guodong Long}
\affiliation{%
  \institution{Australian AI Institute, Faculty of Engineering and IT, University of Technology Sydney}
  \city{Sydney}
  \state{NSW}
  \country{Australia}
}
\email{guodong.long@uts.edu.au}

\author{Chunxu Zhang}
\affiliation{%
  \institution{College of Computer Science and Technology, Jilin University}
  \city{Jilin}
  \country{China}
}
\email{zhangchunxu@jlu.edu.cn}

\author{Honglei Zhang}
\affiliation{%
  \institution{School of Computer Science and Technology, Beijing Jiaotong University}
  \city{Beijing}
  \country{China}
}
\email{honglei.zhang@bjtu.edu.cn}

\author{Jing Jiang}
\affiliation{%
  \institution{Australian AI Institute, Faculty of Engineering and IT, University of Technology Sydney}
  \city{Sydney}
  \state{NSW}
  \country{Australia}
}
\email{jing.jiang@uts.edu.au}

\author{Chengqi Zhang}
\affiliation{%
  \institution{Department of Data Science and Artificial Intelligence, The Hong Kong Polytechnic University}
  \city{Hongkong}
  \country{China}
}
\email{chengqi.zhang@uts.edu.au}


\renewcommand{\shortauthors}{Li et al.}

\begin{abstract}
Federated Recommendation Systems (FRSs) offer a privacy-preserving alternative to traditional centralized approaches by decentralizing data storage. 
However, they face persistent challenges such as data sparsity and heterogeneity, largely due to isolated client environments. Recent advances in Foundation Models (FMs), particularly large language models like ChatGPT, present an opportunity to surmount these issues through powerful, cross-task knowledge transfer. 
In this position paper, we systematically examine the convergence of FRSs and FMs, illustrating how FM-enhanced frameworks can substantially improve client-side personalization, communication efficiency, and server-side aggregation. 
We also delve into pivotal challenges introduced by this integration, including privacy–security trade-offs, non-IID data, and resource constraints in federated setups, and propose prospective research directions in areas such as multimodal recommendation, real-time FM adaptation, and explainable federated reasoning. 
By unifying FRSs with FMs, our position paper provides a forward-looking roadmap for advancing privacy-preserving, high-performance recommendation systems that fully leverage large-scale pre-trained knowledge to enhance local performance. \looseness=-1
\end{abstract}



\received{20 February 2007}
\received[revised]{12 March 2009}
\received[accepted]{5 June 2009}

\maketitle

\section{Introduction}
\label{sec:intro}

In today's digital era, the exponential growth of online information demands recommendation systems (RSs) that can efficiently filter, navigate, and personalize content for individual users. 
Traditional RSs have achieved significant success by tailoring products, content, and services to user preferences~\cite{ko2022survey}. 
However, their heavy reliance on centralized data collection not only raises serious privacy concerns, especially under stringent regulations like GDPR~\cite{voigt2017eu}, but also introduces operational bottlenecks.
To mitigate these issues, Federated Learning (FL) has emerged as a transformative paradigm that enables model training across distributed devices while keeping user data localized~\cite{mcmahan2017communication}. 
By leveraging the computational resources of individual devices, FL alternates between local model updates and global parameter aggregation, giving rise to Federated Recommendation Systems (FRSs) that preserve user privacy~\cite{zhang2021survey}. 
Despite these advantages, FRSs face two critical challenges: 
(a) severe data sparsity: since each client typically contains data from a single user with only a limited set of interactions;
and (b) significant data heterogeneity arising from diverse user behaviors and preferences. 
These challenges often lead to sub-optimal recommendation performance.
In parallel, the recent advent of Foundation Models (FMs) has revolutionized the field of artificial intelligence. 
Language Models such as ChatGPT~\cite{openai2022chatgpt}, vision models like ViT \cite{dosovitskiy2020image}, and multi-modal models like CLIP~\cite{radford2021learning} have demonstrated the power of pre-training on massive, diverse datasets. 
Through techniques such as \textbf{Fine-Tuning}~\cite{zhang2024scaling} and \textbf{Prompting}~\cite{gu2023systematic}, these models can be efficiently adapted to a wide range of downstream tasks, achieving state-of-the-art performance across various domains \cite{dabre2020survey,bommasani2021opportunities,bai2023qwen,oquab2023dinov2,kirillov2023segment}. \looseness=-1

\begin{figure*}[t]
    \centering
    \includegraphics[width=.8\textwidth]{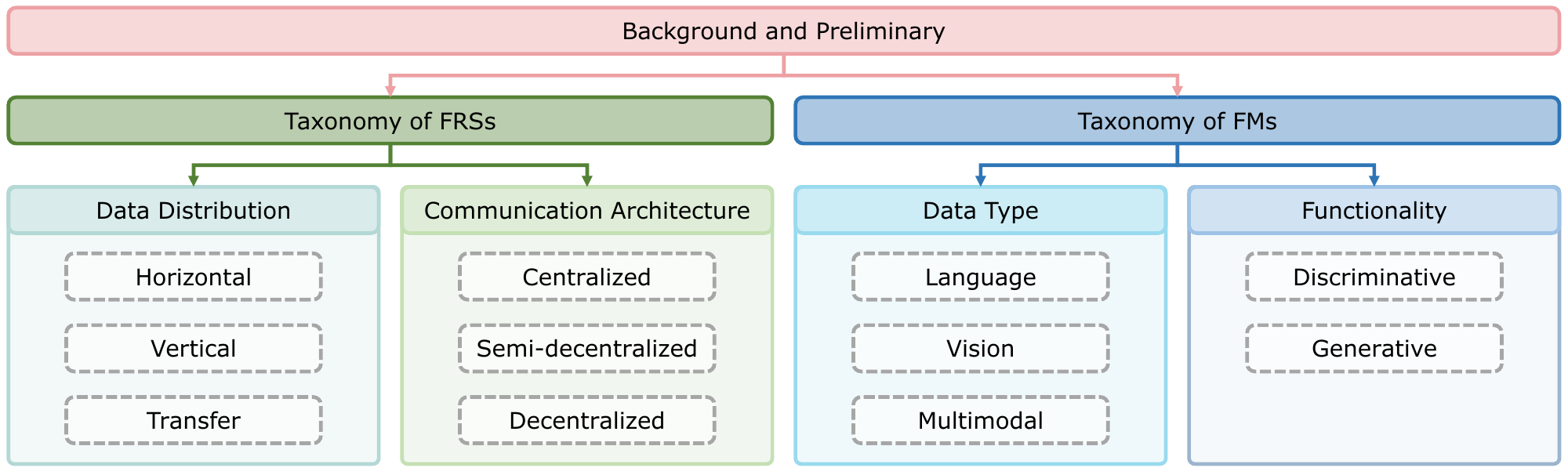}
    \caption{
        The taxonomy of FRS and FM frameworks categorized by their respective core criteria.
    }
    \label{fig:tax}
\end{figure*}

The integration of FMs into FRSs presents a promising avenue to address the challenges of data sparsity and heterogeneity \cite{zhang2024transfr}: 
First, the rich, pre-trained representations provided by FMs can compensate for the limited local data available on each client, thereby enhancing recommendation accuracy. 
Second, the generalization capabilities of FMs help alleviate the cold-start problem by leveraging learned patterns that are broadly applicable to new users and items. 
Third, the transfer learning strengths of FMs allow for rapid adaptation to new recommendation scenarios with minimal additional training. 
Moreover, by reducing the reliance on data sharing, FM-based approaches inherently balance privacy protection with performance, while also mitigating communication overhead \cite{ren2024advances}. 

Despite these advantages, the fusion of FMs and FRSs is still in its early stages. 
Critical issues such as privacy-performance trade-off, communication efficiency, and model fairness remain underexplored. 
This paper is designed to harness the pre-training benefits of FMs while navigating the constraints imposed by federated settings. 
We systematically analyze the challenges associated with this integration and propose future research directions aimed at overcoming these hurdles to promote the development of FRS.

\section{Related Surveys and Contribution}
\label{sec:related}

\looseness=-1
The literature on FRSs has been enriched by several surveys that synthesize methodologies, privacy-preservation techniques, and the challenges inherent in FRS. 
For instance, Yang et al.~\cite{yang2020federated} examine the practical implementation and evaluation of FRSs with a focus on system architectures and algorithmic efficiency, while Alamgir et al.~\cite{alamgir2022federated} provide a comprehensive overview of prevalent techniques, challenges, and prospective research directions. 
Complementary to these works, Javeed et al.~\cite{javeedFederatedLearningbasedPersonalized2024} concentrate on security and privacy issues in personalized RSs, and Sun et al.~\cite{sun2024survey} offer a comparative analysis of current FRS approaches, highlighting both strengths and limitations. 
Collectively, these surveys underscore the critical importance of privacy protection and address the challenges posed by data heterogeneity and model aggregation in FRSs, thereby establishing a solid foundation for further inquiry.

In parallel, the integration of FMs with FL \cite{chen2023federated,zhuang2023foundation,yu2023federated,woisetschlager2024survey,chen2023prompt,ren2024advances} and RSs~\cite{liu2023pre,wu2023survey} has attracted considerable attention. 
However, to the best of our knowledge, no existing work has systematically examined the integration of FMs within FRSs. 
By bridging the gap between FMs and FRSs, our work aim to advance the state-of-the-art in privacy-preserving recommendation technologies in the federated settings, and lay the groundwork for innovative research at the intersection of these two paradigms.

\textbf{Contributions.} Our main contributions are as follows:
\begin{itemize}
    \item We introduce a comprehensive framework for integrating FMs into FRSs, elucidating the core principles and methodologies that enable their seamless fusion.
    \item We demonstrate how the pre-training capabilities of FMs can effectively mitigate issues of data sparsity and heterogeneity to provide better recommendations in federated settings.
    \item We investigate the practical challenges associated with this integration, including privacy--performance trade-offs, communication efficiency, and model generalization, and offer novel insights and potential solutions.
    \item We identify existing research gaps and outline promising future directions to guide subsequent academic inquiry and technological innovation in this emerging field.
\end{itemize}

By elucidating the integration of FMs and FRSs, our work provides a forward-looking roadmap that not only overcomes inherent data and privacy challenges through transferable pre-trained knowledge, but also inspires the next generation of personalized recommendation services in federated settings. \looseness=-1

\section{Background and Preliminary}
\label{sec:bg}

To provide a concise yet profound overview of the current landscape in FRSs and FMs, we summarize the core principles and development trends in each field to lay the groundwork for understanding how their integration can overcome the key challenges in FRSs. 
Moreover, we present detailed taxonomies of both FRSs and FMs as shown in Fig. \ref{fig:tax} based on different criterias, elucidating their diverse architectures and functionalities to further contextualize the potential synergies in this emerging research area.

\subsection{Federated Recommendation Systems}

\begin{figure}[!t]
    \centering
    \includegraphics[width=1.\columnwidth]{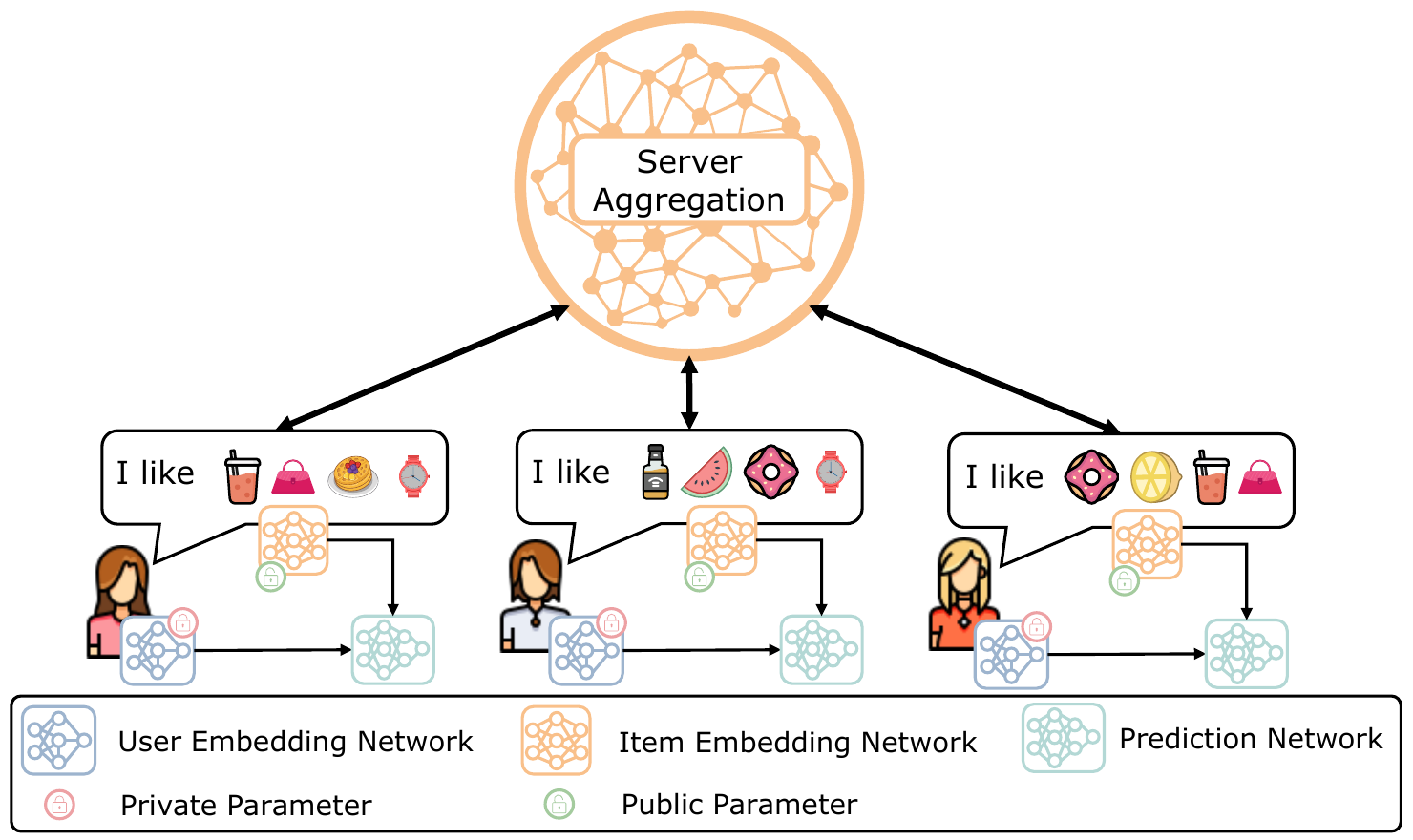}
    \caption{
        A framework for FRS, illustrating client-side local training with private data, server-side global aggregation, and update communication. 
        Local models integrate user and item embeddings with prediction networks, while the server aggregates local updates to enhance the global model, ensuring both privacy and personalization.
    }
    \label{fig:typical_FRS}
\end{figure}

FRSs leverage FL to deliver personalized recommendations while preserving user privacy by ensuring that sensitive data remains on local devices. 
A typical FRS framework, as illustrated in Fig. \ref{fig:typical_FRS}, involves three key stages \cite{sun2024survey}: local model updates, secure transmission of these updates, and global aggregation at central server. 
Though FRSs inherently protect users' data privacy, they still suffer from challenges such as limited and non-IID data at each client, which result in data sparsity and heterogeneity~\cite{yang2020federated}.

\paragraph{Data Distribution Paradigms in FRS.} 
The distribution of data across clients plays a pivotal role in shaping model performance, as the statistical characteristics of local datasets directly impact the model's ability to generalize and capture diverse user behaviors. 
In a \textbf{Horizontal FRS}~\cite{zhang2024federated}, clients share a common feature space while maintaining distinct data samples. 
This scenario is typical when users interact with a shared set of items, yet each client generates personalized data, and the aggregation of their heterogeneous updates contributes to building robust models~\cite{zhang2023dual,li2024federated}. 
In contrast, a \textbf{Vertical FRS} involves clients with overlapping user sets but different feature spaces, enabling the secure integration of complementary data—such as merging financial records with behavioral information—to enrich user profiles and improve recommendation quality~\cite{cao2023privacy,mai2023vertical,wan2023fedpdd}. When individual data sources are extremely limited, a \textbf{Transfer FRS}~\cite{zhuang2020comprehensive,zhang2024transfr} employs transfer learning techniques to share knowledge across domains, thereby mitigating data scarcity and enhancing overall model performance.

\paragraph{Communication Architectures in FRS.} 
The architecture governing client-server communication is pivotal for ensuring both efficiency and security in real-world FRS applications. 
In a \textbf{Centralized FRS}, a central server collects and aggregates client updates, which streamlines the aggregation process and reduces coordination complexity; however, this model inherently creates a single point of failure and concentrates sensitive data, thereby increasing potential privacy risks \cite{zhang2023lightfr,zhang2023dual,li2024federated,zhang2024federated,zhang2024transfr}. 
Alternatively, a \textbf{Semi-decentralized FRS} leverages intermediate nodes such as edge servers to distribute the communication load, thus reducing overall overhead while striking a balance between centralized efficiency and enhanced privacy protection~\cite{qu2023semi}. 
In contrast, a \textbf{Decentralized FRS} employs a peer-to-peer communication model that completely eliminates central coordination, thereby improving privacy by dispersing data and control across the network; however, this approach introduces increased network complexity and necessitates more sophisticated consensus mechanisms to ensure reliable aggregation~\cite{hegedHus2019decentralized,zheng2023decentralized,li2024decentralized}.

\subsection{Foundation Models}

\begin{figure}[t]
    \centering
    \includegraphics[width=1.\columnwidth]{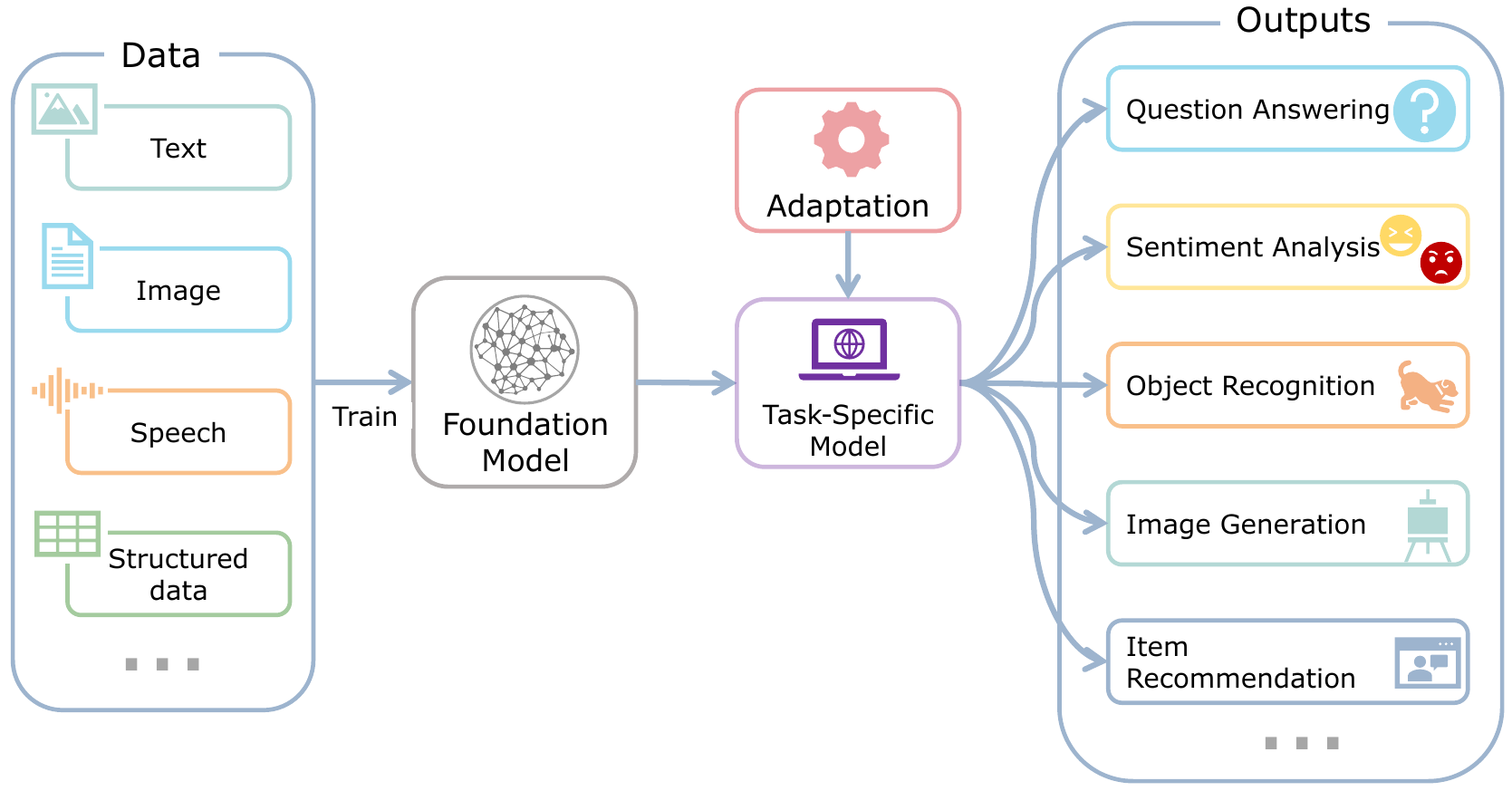}
    \caption{The FM can integrate information contained in data from various modalities during pre-training. The model can then be adapted for a variety of downstream tasks through adapters such as prompting or fine-tuning.}
    \label{fig:fm}
\end{figure}

Recent breakthroughs in hardware, transformer architectures, and large-scale datasets have given rise to Foundation Models, which are capable of transferring learned knowledge across a wide array of tasks \cite{kaplan2020scaling,bommasani2021opportunities,isik2024scaling}. 
As shown in Fig. \ref{fig:fm}, by definition, a foundation model is trained on extensive data via self-supervised learning and can be adapted to diverse downstream applications through techniques such as fine-tuning or prompting~\cite{bommasani2021opportunities}. 
Models like GPT-3 exemplify these systems, showcasing properties of (a) \textit{Emergence}: the spontaneous development of novel capabilities, and (b) \textit{Homogenization}, a unified approach to varied tasks.

\paragraph{Training Data Types in FMs.} The capabilities of FMs are largely determined by the nature of their training data. \textbf{Language FMs} are trained on vast textual corpora and excel in natural language understanding, translation, and text generation~\cite{devlin2018bert,liu2019roberta,brown2020language,touvron2023llama}. \textbf{Vision FMs}, developed using large-scale image datasets, are tailored for visual tasks such as object recognition and segmentation, capturing complex visual patterns~\cite{dosovitskiy2020image,wang2023all,zhang2023recognize,kirillov2023segment}. \textbf{Multi-modal FMs} integrate different data modalities, such as text and images, to support cross-modal applications and deliver richer representations~\cite{radford2021learning,tamkin2021understanding}.

\paragraph{Functional Objectives in FMs.} 
In addition to the data types for training, FMs are also distinguished by their functional objectives. 
\textbf{Discriminative FMs} are engineered to precisely differentiate among various input categories, rendering them highly effective for classification \cite{yao2022prompt} and regression tasks \cite{ali2023evaluating}. Their ability to make fine-grained distinctions has proven valuable in applications ranging from sentiment analysis to predictive modeling~\cite{chen2023robust,ali2023evaluating}. 
Conversely, \textbf{Generative FMs} focus on modeling the underlying data distribution to synthesize new, plausible samples \cite{cao2023comprehensive}. This generative capability, exemplified by models, such as GPT-3~\cite{brown2020language} and DALL·E~\cite{ramesh2021zero}, enables them to produce coherent text \cite{otter2020survey}, create novel images \cite{ho2022cascaded}, and support a wide array of cross-modal applications \cite{cao2023comprehensive}.

\paragraph{Adaptation Techniques for FMs.} 
To customize FMs for specific tasks without retraining the entire model, various adaptation techniques have been developed. \textbf{Prompt-based fine-tuning}~\cite{zhou2022learning} employs learnable prompts to guide model behavior with minimal modifications. 
\textbf{Adapter-based fine-tuning}~\cite{hu2021lora} involves inserting small, trainable modules into the pre-trained network, thereby confining updates to these components while preserving the majority of the original parameters. 
\textbf{External knowledge-based adaptation}~\cite{yang2023language} further enhances performance by incorporating supplementary information, such as domain-specific data or knowledge graphs, into the model’s learning process \cite{pratt2023does}.


\section{Federated Recommendation Systems with Foundation Models}
\label{sec:frs_w_fm}

\begin{figure}[!tb]
    \centering
    \includegraphics[width=1.\columnwidth]{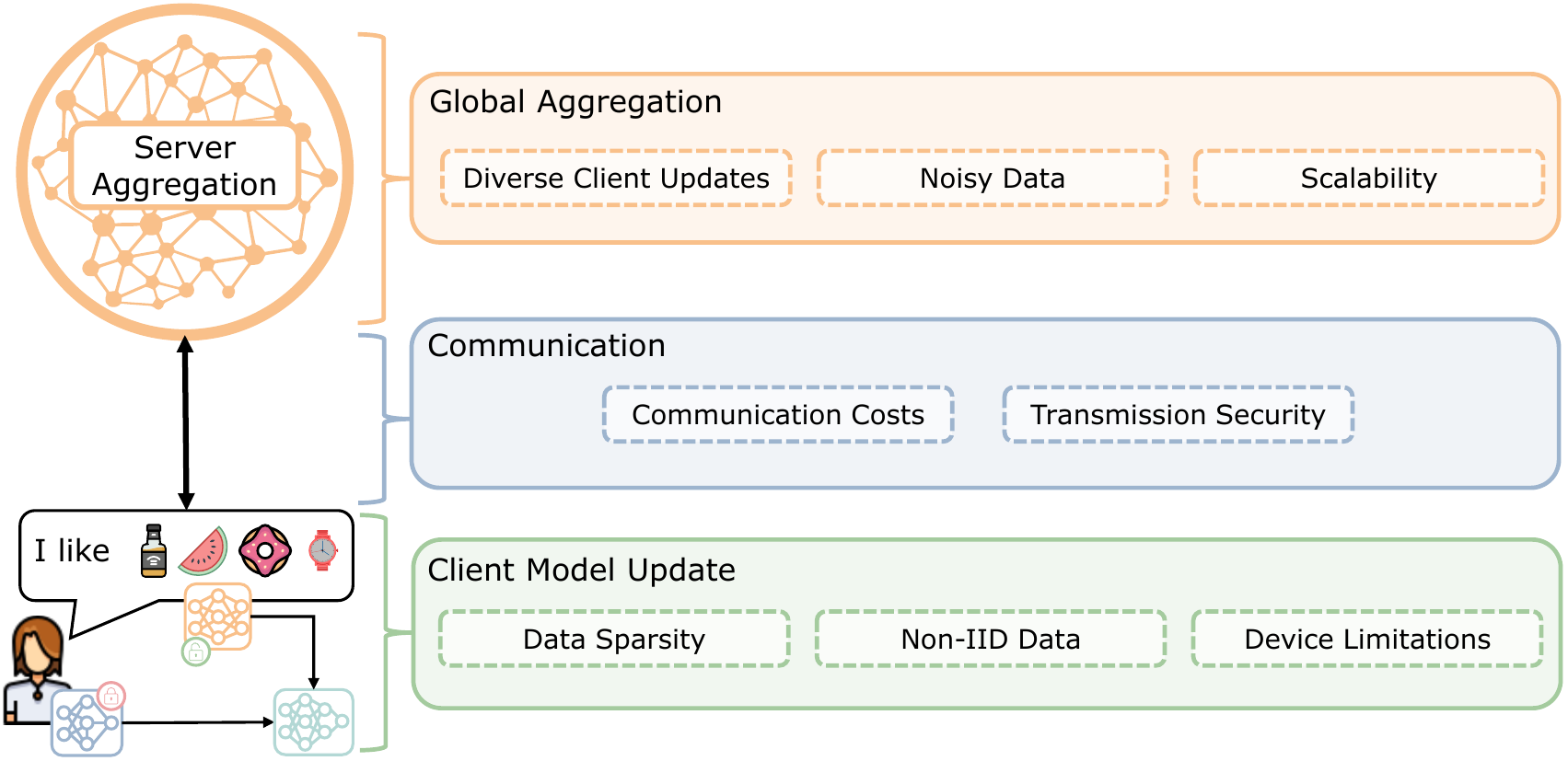}
    \caption{
        Integration framework of FRSs with FMs, illustrating the key challenges across three stages: Client Model Update, Communication, and Global Aggregation.
    }
    \label{fig:frs_fm}
\end{figure}

Though excelling in preserving user privacy by keeping data localized, FRSs are often hindered by data sparsity and heterogeneity due to isolated client environments. 
In contrast, FMs are imbued with rich, transferable knowledge from large-scale pre-training, enabling them to capture complex patterns and semantic nuances across diverse datasets. 
Integrating FMs into FRSs have the ability of offer a promising solution: by leveraging the generalized representations of FMs, client models can be enhanced and global aggregation can be more effectively guided, thus overcoming the limitations imposed by data silos and paving the way for more accurate, scalable, and privacy-aware RSs. 
As illustrated in Fig.~\ref{fig:frs_fm}, the integration framework delineates three pivotal stages: client model update, communication and global aggregation, which we explore in depth in the following sections, emphasizing how FMs can fundamentally address the core challenges of FRSs.

\subsection{Client Model Update}
The client model update stage is inherently challenged by the unique characteristics of decentralized data and the limitations imposed by diverse client devices~\cite{alamgir2022federated}, positing three key issues:

\noindent \textbf{Data Sparsity}:  
Due to stringent privacy requirements, user data, including interaction histories, personal profiles, and other sensitive information, remains confined to local devices, thereby forming isolated data silos \cite{yang2020federated}. 
While this decentralized approach is crucial for complying with privacy regulations and enhancing user trust, it also means that each client holds only a limited portion of the overall dataset \cite{li2023federated}. 
As a result, the inherent data sparsity significantly hampers the model’s ability to learn robust, generalized representations, ultimately posing a major challenge to achieving optimal performance in federated recommendation scenarios \cite{zhang2024transfr}.

\noindent \textbf{Non-IID Data}:  
Since each client independently trains the model on its localized data, the aggregated data distribution is inherently non-independent and identically distributed (non-IID) \cite{sun2024survey}. 
This heterogeneity, reflecting the diverse preferences, usage patterns, and behaviors of individual users, enhances personalization by capturing unique user characteristics \cite{zhang2023dual}. 
However, it also complicates the training process, as the global model must reconcile conflicting updates and varying feature distributions, often leading to slower convergence and potential degradation in performance.

\noindent \textbf{Device Limitations}:  
In FRS, clients typically operate on consumer-grade devices such as smartphones, which are constrained by limited computational resources, and often face unstable connectivity \cite{alamgir2022federated}. 
Consequently, it is essential to minimize the computational burden of local model updates to ensure that training tasks are executed efficiently without depleting device resources. 
Furthermore, reducing the volume of data exchanged is critical to accommodate fluctuating network conditions, lower latency, and maintain an overall responsive user experience in real-world deployments.

FMs, pre-trained on vast and diverse datasets, offer robust representational capabilities that capture intricate semantic patterns and contextual nuances~\cite{bommasani2021opportunities}. 
The inherent adaptability enables FMs to be efficiently fine-tuned to local conditions, thereby bridging the gap between globally learned knowledge and client-specific data, which yields several advantages in the client update phase : 

\noindent \textbf{Addressing Data Sparsity}: 
By leveraging transfer learning, FMs can transfer broad, generalized knowledge from large corpora to enhance local, sparse datasets~\cite{zeng2024federated}. 
Fine-tuning an FM on limited client data not only enables effective adaptation to individual user behaviors, thereby improving recommendation accuracy~\cite{wu2023survey}, but also bridges the gap between global pre-training and local contextual nuances. 
Moreover, this localized fine-tuning ensures that users' sensitive data remains on their device, thereby upholding stringent privacy standards in federated settings~\cite{rame2023model}.

\noindent \textbf{Handling Non-IID Data}: FMs possess powerful semantic understanding that allows them to capture complex patterns from non-IID data. Their ability to interpret diverse inputs, such as search queries~\cite{liu2021pre}, user comments~\cite{liu2023pre}, and other textual signals~\cite{geng2022recommendation}, enables a more personalized recommendation process in FRS, effectively mitigating the challenges of data heterogeneity~\cite{li2024feddae,tang2024higpt}.

\noindent \textbf{Adapting to Device Limitations}: Although fine-tuning FMs can be computationally demanding, recent advances in lightweight fine-tuning techniques and model compression have significantly reduced the associated overhead \cite{han2024parameter,chen2024percept}. 
These innovations not only enable efficient deployment of complex FMs on resource-constrained devices but also facilitate real-time personalization and faster inference. 
Consequently, even in federated environments with limited computational resources, these methods ensure that the enhanced performance of FMs can be fully leveraged to deliver high-quality, personalized recommendations in federated settings.

Overall, by integrating FMs during the client update stage, FRSs can effectively mitigate inherent limitations such as data sparsity, non-IID distributions, and device constraints. 
This integration leverages the deep, pre-trained representations of FMs to enrich local models, enabling them to learn more robustly from limited and heterogeneous data. 
Consequently, local model performance is significantly enhanced by integrating FMs, leading to more accurate, personalized recommendations across the federated network.

\subsection{Communication}

The communication phase in FRS, transferring model updates from clients to the central server, is crucial for system performance and privacy, yet it faces two major challenges~\cite{sun2024survey}:

\noindent \textbf{High Communication Costs}:  
In FRS, the transmission of high-dimensional model parameters, gradients, or other statistical summaries from numerous clients can impose substantial communication overhead. 
This challenge becomes even more pronounced in large-scale systems, where constrained bandwidth and elevated latency further exacerbate data exchange inefficiencies~\cite{li2023federated}.

\noindent \textbf{Transmission Security}:  
Model updates often contain sensitive information that, if intercepted, could compromise user privacy \cite{chai2020secure}. 
Although encryption and secure aggregation techniques \cite{sun2024survey} have been developed to safeguard user data, their deployment typically increases additional computational costs and communication complexity, thereby potentially impacting overall system performance.

FMs offer promising solutions by leveraging advanced representation and compression capabilities. Pre-trained on vast, diverse datasets, they acquire rich prior knowledge that can be rapidly adapted to specific client data through fine-tuning~\cite{bommasani2021opportunities}. This enables FMs to learn general feature representations and adapt to varying data distributions, significantly enhancing local model performance when applied during the client update phase in FRS.
Specifically:

\noindent \textbf{Addressing Data Sparsity}: 
FMs can apply the knowledge learned from large corpora to local data through transfer learning \cite{zeng2024federated}. 
For instance, a pre-trained FM can be effectively fine-tuned on a small amount of local data at the client to adapt to specific user behaviors, thereby achieving good performance in downstream tasks such as providing more accurate recommendations \cite{wu2023survey,yuan2024fellas}. 
Furthermore, by fine-tuning the base model locally, user sensitive data does not need to leave the device, protecting user privacy \cite{rame2023model}. 
Thus, even with limited data in FRS, clients can achieve better model performance, alleviating the challenges brought by data sparsity.

\noindent \textbf{Handling Non-IID Data}: The strong representational capabilities of FMs enable them to capture the complex user preferences and behaviors. 
For example, FMs possess powerful semantic understanding abilities, allowing them to better interpret user search queries \cite{lu2020twinbert,liu2021pre}, comments \cite{liu2023pre}, and other textual data \cite{geng2022recommendation}, thereby providing recommendations that align more closely with user needs. 
Moreover, FMs can learn complex patterns from non-IID data \cite{han2021pre,li2024feddae,tang2024higpt}, enhancing the model's adaptability to different user preferences and ensuring that recommendation results stay consistent with the user's current interests and needs. 
Applying FMs during the client update phase can achieve a higher level of personalized recommendation, effectively addressing data heterogeneity in FRS.

\noindent \textbf{Adapting to Device Limitations}: 
Though fine-tuning and updating FMs typically require significant computational resources, optimizing the model architecture and employing lightweight fine-tuning techniques can reduce the computational burden on devices \cite{han2024parameter}. 
The knowledge transfer capability of FMs allows for efficient local fine-tuning under limited computational resources, thereby achieving good recommendation performance even in resource-constrained environments \cite{chen2024percept}. 
Additionally, model compressions \cite{zhu2023survey} can be used to reduce the model size, thereby improve communication efficiency to suit the limitations of user devices.

\subsection{Global Aggregation}

The global aggregation phase of FRSs is responsible for synthesizing the diverse and heterogeneous updates received from clients into a cohesive global model \cite{alamgir2022federated}. 
This stage is pivotal for ensuring overall effectiveness of personalized recommendations, yet it must contend with several challenges in federated settings \cite{sun2024survey}:

\noindent \textbf{Diverse Client Updates}:  
Central servers are tasked with integrating updates uploaded by clients, each reflecting distinct data distributions and unique user behaviors \cite{mcmahan2017communication}. 
This diversity introduces significant challenges in constructing a coherent global model that accurately captures the underlying trends across all clients. 
Therefore, effective aggregation on the server is crucial to reconcile these variations and ensure that the final model maintains high quality and recommendation accuracy while privacy preserving.

\noindent \textbf{Noisy Data}:  
Client updates may include noisy or anomalous data that can significantly degrade the performance of the global model  \cite{zhang2024hn3s}. 
Robust aggregation methods \cite{zhang2024federated} are therefore essential to identify and filter out these inconsistencies from clients, while preserving the valuable information contained in accurate updates.

\noindent \textbf{Scalability}:  
As the number of clients and the volume of data in FRS continue to surge, the aggregation process faces mounting scalability challenges \cite{sun2024survey}. 
The system must efficiently consolidate updates from a vast and diverse array of sources, all while contending with increased computational demands and potential communication delays \cite{chai2020secure}. 
This rapid expansion intensifies the complexity of synchronizing model updates, posing significant obstacles to maintaining timely and efficient global model convergence in FRS.

Traditional methods such as the weighted averaging approach used in FedAvg~\cite{mcmahan2017communication} do not fully exploit the contextual richness of client updates. 
In contrast, FMs can enhance the effectiveness of the global aggregation in FRS through several innovative strategies:

\noindent \textbf{Context-Aware Aggregation}: 
FMs can leverage their powerful representation capabilities to analyze the structural and statistical characteristics of client updates \cite{bommasani2021opportunities}, such as update frequency, magnitude, and temporal trends. 
By extracting and interpreting these multifaceted features, FMs enable the assignment of contextually appropriate weights to each update~\cite{yu2023federated}. 
This nuanced weighting helps align the aggregation process with the inherent system-wide patterns, thereby enhancing the fidelity of the global model.

\noindent \textbf{Dynamic Weighting}: 
FMs facilitate dynamic adjustments of update weights based on their relevance and contribution to global parameters \cite{zhou2024comprehensive}. 
For example, client updates from under-represented data distributions may be assigned higher weights, ensuring that the aggregated model remains comprehensive and robust by capturing a wider range of user behaviors and mitigating inherent data imbalances~\cite{han2024parameter}. This mechanism enables the global model to adapt continuously to the evolving patterns present in client data.

\noindent \textbf{Knowledge-Based Aggregation}: By harnessing their extensive pre-trained knowledge, FMs can infer latent relationships and underlying patterns from client updates that may not be immediately evident in the raw data. 
This capability allows them to effectively compensate for sparse or inconsistent inputs by contextualizing limited data within a broader semantic framework. 
Such a knowledge-based approach enriches the aggregation process, particularly in cases where certain clients provide insufficient data, ultimately contributing to a more robust and comprehensive global model\cite{chen2024large}.

\noindent \textbf{Anomaly Detection and Handling}: FMs excel at detecting anomalies in client updates by identifying deviations from typical patterns and expected trends \cite{xu2024customizing}. 
Leveraging the deep semantic understanding, FMs can pinpoint outlier updates that may signal errors or inconsistencies in the data. 
By dynamically down-weighting or excluding these outliers, FMs help preserve the stability and robustness of the global model, ensuring that unpredictable fluctuations in client data do not compromise overall performance~\cite{liu2023pre}.

Overall, the integration of FMs into the global aggregation phase empowers FRSs to intelligently synthesize diverse client updates by leveraging deep, contextual insights to reconcile discrepancies across heterogeneous data sources. 
By incorporating advanced techniques—such as context-aware analysis that captures nuanced data patterns, dynamic weighting that adjusts contributions based on relevance, knowledge-based inference to compensate for sparse inputs, and robust anomaly handling to filter out inconsistencies, FMs significantly enhance the quality, robustness, and adaptability of the aggregated model. 
This positions FMs as a transformative asset for developing personalized, scalable, and privacy-preserving recommendation systems in federated environments. 
Continued research in this area promises further innovations, driving the evolution of FRSs toward more intelligent, secure, and resilient systems capable of meeting real-world demands.

\section{Key Challenges and Strategic Solutions}
\label{sec:challenge}

\begin{figure}[!tb]
    \centering
    \includegraphics[width=.75\columnwidth]{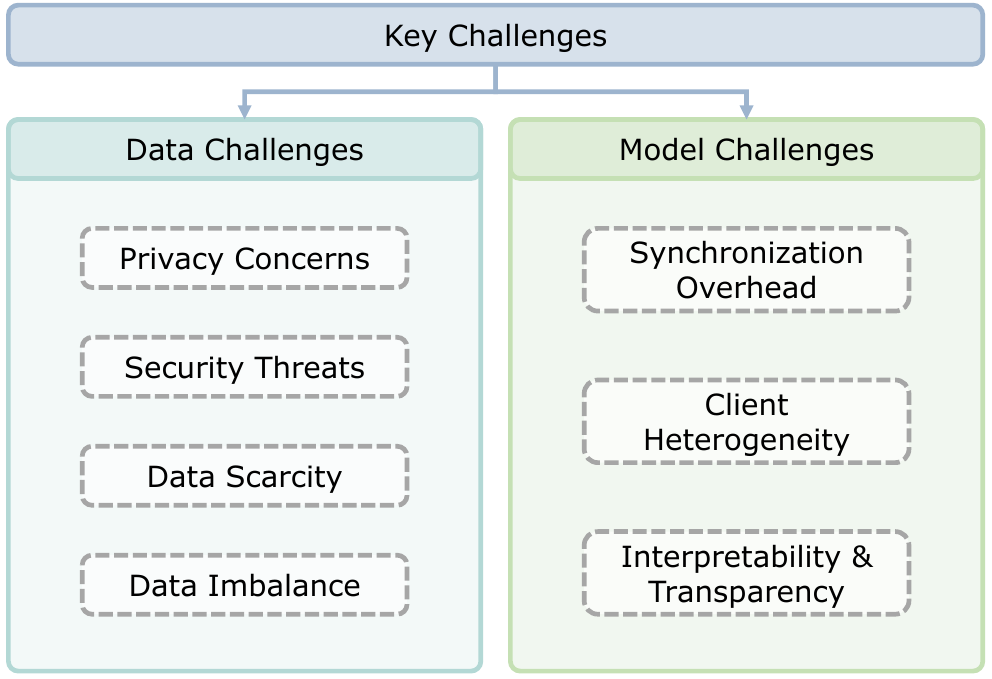}
    \caption{
        Challenges in integrating FMs into FRSs, divided into data and model challenges, highlighting areas requiring strategic solutions for enhanced personalization and privacy.
        }
    \label{fig:chal}
 \end{figure}

Integrating FMs into FRSs opens up unprecedented opportunities to enhance personalized recommendations while rigorously maintaining user privacy. 
However, as illustrated in Fig.~\ref{fig:chal}, this integration also introduces significant challenges across multiple dimensions. 
These challenges include managing non-IID data distributions, addressing limited computational resources on client, overcoming communication bottlenecks, and ensuring robust and fair aggregation of diverse client updates. 
A comprehensive understanding of these issues, and the development of targeted strategies to address them, is essential for advancing FRSs. 
In this section, we delve deeply into these key challenges and outline potential strategic solutions that can pave the way for more intelligent and secure FRSs. \looseness=-1

\subsection{Data Challenges and Mitigation Strategies}

\subsubsection{Privacy Concerns}
Privacy aims to prevent unauthorized access or misuse of data, especially during model training and data handling, to protect user identities and personal details \cite{hasan2023security}. 
In a federated setting, safeguarding user privacy is of utmost importance. 
Privacy concerns focus on preventing unintended exposure or misuse of personal data, ensuring that sensitive information remains confidential \cite{citron2022privacy}. 
While training and recommendations must occur without revealing personal data, FMs, such as GPT-3 \cite{brown2020language}, may memorize and reproduce training data, potentially leaking sensitive information \cite{chen2023hide}. 
Additionally, if generated data closely resembles original data, user privacy risks may arise \cite{hasan2023security}.

To mitigate these privacy concerns, differential privacy \cite{minto2021stronger} can be employed by adding noise during model training to reduce data leakage risks while preserving model performance. 
Homomorphic encryption \cite{acar2018survey} allows training on encrypted data, effectively preventing data theft during transmission and processing. 
Moreover, machine unlearning techniques \cite{bourtoule2021machine} can be utilized to remove specific user data from the FMs, ensuring compliance with privacy regulations like GDPR.
An emerging approach, known as privacy rewrite, transforms sensitive data into anonymized formats before transmission. Frameworks like HaS \cite{chen2023hide} exemplify this by anonymizing private entities locally and reconstructing them after processing. This method can reduce privacy risks, preserves data utility, and minimizes computational overhead in FRSs. 

\subsubsection{Security Threats}
Security threats involve protecting data from malicious attacks or breaches~\cite{hasan2023security}. 
While privacy focuses on controlling who can access and use the data, security emphasizes preserving data integrity and defending against cyber threats such as unauthorized data manipulation and theft~\cite{nasr2019comprehensive}. 
In the integration of FMs into FRSs, ensuring robust data security is critical. 
Participants may face various attacks, including (a) Member Inference Attacks~\cite{nasr2019comprehensive}, where adversaries aim to deduce sensitive information about individual clients; (b) Data Reconstruction Attacks~\cite{lyu2021novel}, which attempt to recover original data from model updates; and (c) Poisoning Attacks~\cite{tolpegin2020data}, wherein malicious data is injected to corrupt the model. 
These threats can compromise both the integrity of the global model and the quality of the underlying data, highlighting the urgent need for comprehensive security measures in FM-enhanced federated environments to preserve user privacy.

To ensure data integrity and confidentiality during transmission, secure multi-party computation (SMPC) techniques~\cite{byrd2020differentially} can be employed to distribute computational tasks among multiple participants, ensuring that no single party gains access to the complete dataset. This collaborative approach mitigates the risk of data breaches during processing. Additionally, blockchain technology~\cite{hai2022bvflemr,guo2023decentralized} can be utilized to record data access and operations in an immutable ledger, thereby enhancing data integrity, traceability, and transparency. Through consensus mechanisms and verifiable records, blockchain further strengthens the overall security framework, fostering trust among participants in federated systems.

\subsubsection{Data Scarcity}
Despite the promise of FMs to alleviate data sparsity, FRSs still often confront significant data scarcity~\cite{alamgir2022federated}, especially in federated environments where participants generate only limited interaction data, e.g., the users may only visit a small amount of items. 
This scarcity can severely impair model performance, as the insufficient volume of user interactions hinders the development of robust and effective recommendation models~\cite{li2024personalized}.

Data augmentation techniques~\cite{mumuni2022data} can be employed to generate diverse and realistic training samples, thereby effectively expanding the dataset and capturing a broader range of user behaviors. 
In parallel, leveraging knowledge transfer from rich datasets in other domains~\cite{chen2024percept} provides an additional means to alleviate data sparsity, ultimately enhancing overall model performance by infusing external contextual insights. 
Furthermore, when employing generative FMs~\cite{rossi2024augmenting} for synthesizing new data, it is imperative to implement robust quality control mechanisms to ensure that the synthetic data meets stringent quality standards and does not introduce biases that could adversely affect the model on the client.

\subsubsection{Data Imbalance}
In federated settings, significant differences in data size and distribution exist across clients, as the users' behaviors diverse, leading to pronounced data imbalance~\cite{duan2020self,wang2021addressing,luo2022towards}. 
This imbalance is often driven by long-tail distributions of item labels and user behaviors, where a few clients contribute the bulk of the data while many others provide only sparse interactions. 
Such disparities hinder model training effectiveness by skewing the global model towards data-rich clients, ultimately limiting its ability to generalize across the full spectrum of user behaviors.

To address sample imbalance, various strategies can be employed. 
For instance, resampling techniques such as oversampling minority classes or under-sampling majority classes~\cite{he2009learning} can help re-balance the dataset by either replicating underrepresented samples or reducing the dominance of abundant ones. 
Alternatively, weighted loss functions~\cite{li2021autobalance,duan2020self} can be applied to assign greater importance to minority class samples during training, thereby enhancing the model’s sensitivity to underrepresented data without altering the intrinsic data distribution on each client.

\subsection{Model Challenges and Strategic Solutions}

\subsubsection{Synchronization Overhead}
In FRS, the need for frequent synchronization between clients and the central server can result in high communication costs and increased system complexity. 
This issue is particularly exacerbated when dealing with FMs that contain extensive parameters, as each synchronization round may involve transferring a significant amount of data. 
Such overhead not only strains network resources but also complicates the coordination and consistency of model updates across distributed devices~\cite{bommasani2021opportunities}.

Gradient compression techniques~\cite{lin2017deep} can significantly reduce the volume of data transmitted during synchronization by compressing or eliminating redundant gradient information, thereby lowering overall communication costs. Moreover, asynchronous update strategies~\cite{chen2019communication} allow clients to update their models independently rather than waiting for synchronous rounds, enabling periodic global aggregation. 
This decoupling of local updates from global synchronization not only reduces idle times but also enhances communication efficiency, particularly when managing FMs with extensive parameters and large amount of clients.

\subsubsection{Client Heterogeneity}
Arising from variations in model types, sizes, and architectures, client heterogeneity creates significant challenges when deploying a unified FM in federated settings~\cite{zhu2018knowledge}. The diverse computational capacities, memory constraints, and network conditions across clients further complicate the synchronization and consistent performance of a single FM. Consequently, achieving uniform model behavior across such heterogeneous environments demands innovative adaptive strategies in FRSs.

Adaptive training algorithms~\cite{tian2019contrastive,zhang2021parameterized} that dynamically adjust model parameters based on each client's computational capabilities and data characteristics show great promise in addressing the challenge of client heterogeneity in FRS. These methods tailor the training process to accommodate variations in device performance and localized data distributions, thereby optimizing learning efficiency on a per-client basis. 
Moreover, inter-client knowledge transfer~\cite{zhang2024transfr} can further mitigate disparities by sharing learned representations across clients, ultimately enhancing the global model's performance and robustness while preserving privacy.

\subsubsection{Interpretability and Transparency}
FMs are often perceived as black-box models, which obscures their internal workings and hampers interpretability. 
This lack of transparency not only complicates the diagnosis of errors and the understanding of decision-making processes but also raises significant concerns regarding trust and regulatory compliance~\cite{ren2024advances}. 
The opaqueness of these models can undermine stakeholder confidence, making it imperative to develop methods that enhance explainability and ensure that model behavior aligns with ethical and legal standards in FRSs.

Explainable AI techniques~\cite{gunning2019xai}, including attention mechanisms~\cite{mohankumar2020towards} and feature importance analysis~\cite{carletti2019explainable}, offer critical insights into the internal decision-making processes of models, thereby enhancing transparency within FRSs. 
Furthermore, the use of generative FMs~\cite{schneider2024explainable} enables the production of coherent natural language explanations for recommendations, which not only elucidate the underlying rationale but also foster user trust and acceptance.

\begin{figure}[!tb]
    \centering
    \includegraphics[width=.85\columnwidth]{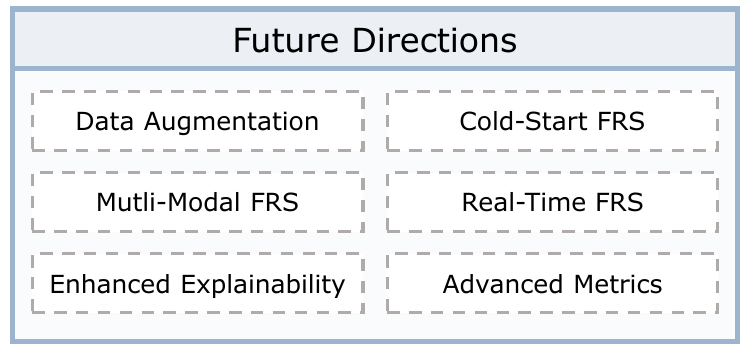}
    \caption{
        Key future directions for integrating FMs into FRSs aim to address emerging challenges and opportunities, focusing on enhancing recommendation performance, safeguarding user privacy, and improving system adaptability.
        }
    \label{fig:dir}
 \end{figure}

\section{Future Directions}
\label{sec:dir}

As shown in Fig.~\ref{fig:dir}, The convergence of FMs and FRSs opens a promising yet challenging frontier. 
By harnessing the powerful representation capabilities of FMs alongside the privacy-preserving strengths of FRSs, researchers can transcend longstanding obstacles such as data sparsity and heterogeneity, while simultaneously boosting recommendation performance. 
In this section, we outline several critical future directions that not only exploit the intrinsic advantages of FMs but also address the new challenges they introduce. 
These directions serve as a comprehensive roadmap for advancing research and practical applications toward more robust and privacy-preserving recommendation services.

\subsection{Data Augmentation}
Data scarcity is a big challenge in FRSs due to stringent privacy constraints that confine user data to local devices \cite{sun2024survey}. 
Generative FMs offer a compelling solution by synthesizing realistic user interaction data to augment training sets~\cite{yang2022image}. 
By generating virtual interaction records and detailed item descriptions, these models can enrich sparse user profiles, thereby enhancing recommendation accuracy and personalization~\cite{yan2019differentiated}. 
Nonetheless, the quality of synthetic data must be rigorously validated to mitigate potential biases and noise~\cite{maharana2022review}. 
Future research should focus on developing robust methods for generating diverse and high-quality synthetic data, tailored to cover a wide range of user behaviors and scenarios.

\subsection{Cold-Start Recommendation}
The cold-start problem, stemming from insufficient historical interaction data for new users or items, remains a significant hurdle in FRSs~\cite{schein2002methods}. 
Pre-trained FMs, endowed with rich semantic knowledge from vast textual corpora, can generate high-quality representations for both users and items, thereby enabling zero-shot and few-shot learning approaches~\cite{gong2023unified,zhang2024federated}. 
This capability not only mitigates the cold-start issue but also facilitates a smoother adaptation to novel scenarios. 
However, ensuring robust privacy in federated settings and effectively transferring knowledge across domains without compromising model integrity remain challenging. 
Future work must advance privacy-preserving mechanisms and efficient knowledge transfer techniques to further bolster cold-start recommendations while maintaining stringent privacy standards.

\subsection{Multi-Modal Recommendation}
One of the standout strengths of FMs is their ability to process and integrate multiple data modalities, such as text, images, audio, and video, to construct richer and more nuanced user profiles~\cite{bommasani2021opportunities}. 
Incorporating multi-modal data into FRSs can lead to significantly enhanced personalization and recommendation quality~\cite{wang2021survey}. 
However, the inherent heterogeneity of different data types poses a considerable challenge in terms of unified representation. 
Future research should focus on developing sophisticated methods to map diverse modalities into a common latent space, while concurrently designing robust, privacy-preserving protocols to safeguard sensitive multi-modal information~\cite{li2024personalized,feng2024robust,wu2024towards}.

\subsection{Real-Time Recommendations}
Real-time RSs are crucial for dynamically adapting to evolving user behaviors and contextual cues~\cite{huang2015tencentrec}. 
FMs can enhance the accuracy and relevance of these recommendations by leveraging their advanced contextual understanding to process user queries and item descriptions in real time. 
Nevertheless, the high computational demands of FMs may introduce latency, adversely affecting the user experience. 
Future research should prioritize the development of model compression and acceleration techniques, such as knowledge distillation~\cite{kang2022personalized} and pruning~\cite{beel2019data}, to reduce computational complexity. 
Additionally, efficient context management strategies, e.g., sliding window approaches \cite{huang2015tencentrec}, should be explored to optimize the handling of continuous user behavior streams.

\subsection{Enhanced Explainability}
Explainability is pivotal for fostering user trust and satisfaction in recommendation systems \cite{hada2021rexplug,geng2022path}. Language FMs, pre-trained on extensive textual datasets, are well-equipped to generate coherent, natural language explanations that elucidate the rationale behind recommendations~\cite{liu2023chatgpt}. However, producing these detailed explanations incurs substantial computational costs, and there is a significant risk of perpetuating biases embedded in the pre-training data~\cite{bommasani2021opportunities}. Future research should focus on balancing the trade-off between explanation quality and computational efficiency through advanced model optimization and robust debiasing techniques. Moreover, incorporating user feedback into the explanation generation process can further refine and enhance the fairness and clarity of recommendations, while also ensuring that privacy is preserved.

\subsection{Advanced Metrics}
Evaluating FRSs integrated with FMs necessitates the development of advanced metrics that extend beyond conventional measures such as rating prediction and item ranking~\cite{wu2023survey}. 
Given the generative and explanatory capabilities of FMs, novel evaluation frameworks must capture additional dimensions, including diversity, fairness, contextual relevance, and overall user satisfaction, to provide a comprehensive assessment of these hybrid systems. 
Such holistic criteria offers deeper insights into model performance, and reveal latent trade-offs that guide further optimization and innovation \cite{wu2024survey}. \looseness=-1

To conclude, the integration of FMs into FRSs heralds a transformative shift toward more intelligent, adaptable, and privacy-conscious recommendation services. 
While the potential benefits are substantial, addressing the associated challenges, ranging from data quality and computational efficiency to privacy and fairness, demands a concerted, multidisciplinary research effort. 
The future directions outlined above offer a strategic roadmap for pioneering advancements in this emerging field, paving the way for next-generation RSs that are robust, scalable, and truly user-centric.



\section{Conclusion}
\label{sec:conclusion}

This paper has explored the integration of Federated Recommendation Systems with Foundation Models, demonstrating that leveraging pre-trained knowledge through lightweight adaptation effectively addresses challenges such as data sparsity, non-IID distributions, and device limitations while preserving user privacy. 
By enhancing both local model performance and global aggregation, this synergy mitigates key issues like privacy-performance trade-offs and communication bottlenecks. 
Looking forward, promising research avenues include generative data augmentation, cold-start mitigation, multi-modal fusion, and real-time adaptation, all of which are pivotal for developing robust, scalable, and user-centric recommendation systems. 
In essence, the fusion of FRSs and FMs offers a transformative pathway toward next-generation recommendation systems that are both highly effective and intrinsically privacy-preserving in the federated settings.

\bibliographystyle{ACM-Reference-Format}
\bibliography{survey}

\end{document}